\documentclass{aa}  

\usepackage{graphicx}
\usepackage{txfonts}

\usepackage{gensymb}
\usepackage{natbib}
\usepackage{subfig}
\usepackage{multirow}
\usepackage{mathtools}
\usepackage{array}
\usepackage{graphicx}
\usepackage{hyperref}
\bibpunct{(}{)}{;}{a}{}{,}
\usepackage{nicefrac}
\usepackage{pdflscape}
\usepackage{rotating}

\def\degr{\hbox{$^\circ$}}
\def\arcmin{\hbox{$^\prime$}}
\def\arcsec{\hbox{$^{\prime\prime}$}}
\newcommand{\microns}{$\mu$m}
\newcommand{\herschel}{\emph{Herschel}}
\newcommand{\spitzer}{\emph{Spitzer}}
\newcommand{\higal}{Hi-GAL }

\begin{document} 

\title{The Initial Conditions of Stellar Protocluster Formation}

\subtitle{III. The \herschel\ counterparts of the \spitzer\ Dark Cloud catalogue}

\author{N. Peretto\inst{1}, C. Lenfestey\inst{2}, G. A. Fuller\inst{2}, A. Traficante\inst{2}, S. Molinari\inst{3}, M. A. Thompson\inst{4}, D. Ward-Thompson\inst{5}}

\institute{School of Physics \& Astronomy, Cardiff University, Queen's building, The parade, Cardiff, CF24 3AA, UK
\and{Jodrell Bank Centre for Astrophysics, Alan Turing building, School of Physics \& Astronomy, The University of Manchester, Oxford Road, Manchester M13 9PL, UK}
\and{IAPS-Istituto di Astrofisica e Planetologia Spaziali, Via Fosso del Cavaliere 100, I-00133 Roma, Italy}
\and{Centre for Astrophysics Research, University of Hertfordshire, College Lane, Hatfield, Herts AL10 9AB, UK}
\and{Jeremiah Horrocks Institute, University of Central Lancashire, Preston, Lancashire PR1 2HE, UK}
}

\abstract {Galactic plane surveys of pristine molecular clouds are key for
  establishing a Galactic-scale view of the earliest stages of star formation. For this reason
  Peretto \& Fuller (2009; hereafter PF09) built an unbiased sample of Infrared Dark Clouds
  (IRDCs) in the $10\degr < |l| < 65\degr, |b|<1\degr$ region of the Galactic
  plane using  \spitzer\ 8\microns\ extinction. However, in absorption studies,  intrinsic fluctuations in the mid-infrared background can be mis-interpreted as foreground clouds. }
{The main goal of the study presented here is to disentangle real clouds in the \spitzer\ Dark Cloud (SDC)  catalogue from artefacts due to fluctuations in the  mid-infrared background.} 
 {We constructed H$_2$ column density maps at $\sim18\arcsec$ resolution  using the 160\microns\ and 250\microns\ data from the \herschel\ Galactic plane survey
  Hi-GAL.  We also developed an automated detection scheme that confirms the existence of a SDC through its association with a peak on these  \herschel\ column density maps. Detection simulations, along with visual inspection of a small sub-sample of SDCs, have been performed to get better insight into the limitations of our  automated identification scheme. } 
{ 
 Our analysis shows that   $76(\pm19)\%$ of the catalogued SDCs are real. This fraction drops to $55(\pm12)\%$ for clouds  with angular diameters larger than $\sim1$ arcminute.  The contamination of the PF09 catalogue by large spurious sources reflect the large uncertainties associated to the construction of the 8\microns\ background emission, a key stage towards the identification of SDCs. A comparison of the \herschel\ confirmed SDC sample with the BGPS and ATLASGAL samples shows that SDCs probe a unique range of cloud properties, reaching  down to more compact and lower column density clouds than any of these two (sub-)millimetre Galactic plane surveys.}
  {Even though  about half of the large SDCs are revealed to be spurious sources, the vast majority of the catalogued SDCs do have a \herschel\ counterpart. The \herschel\ confirmed sample of SDCs offers a unique opportunity to study the earliest stages of both low- and high-mass star formation across the Galaxy.} 


\titlerunning{The \herschel\ counterparts of \spitzer\ dark clouds}
\authorrunning{N. Peretto et al.}

\maketitle

\begin{figure*}[ht]
\centering
\includegraphics[width=17cm]{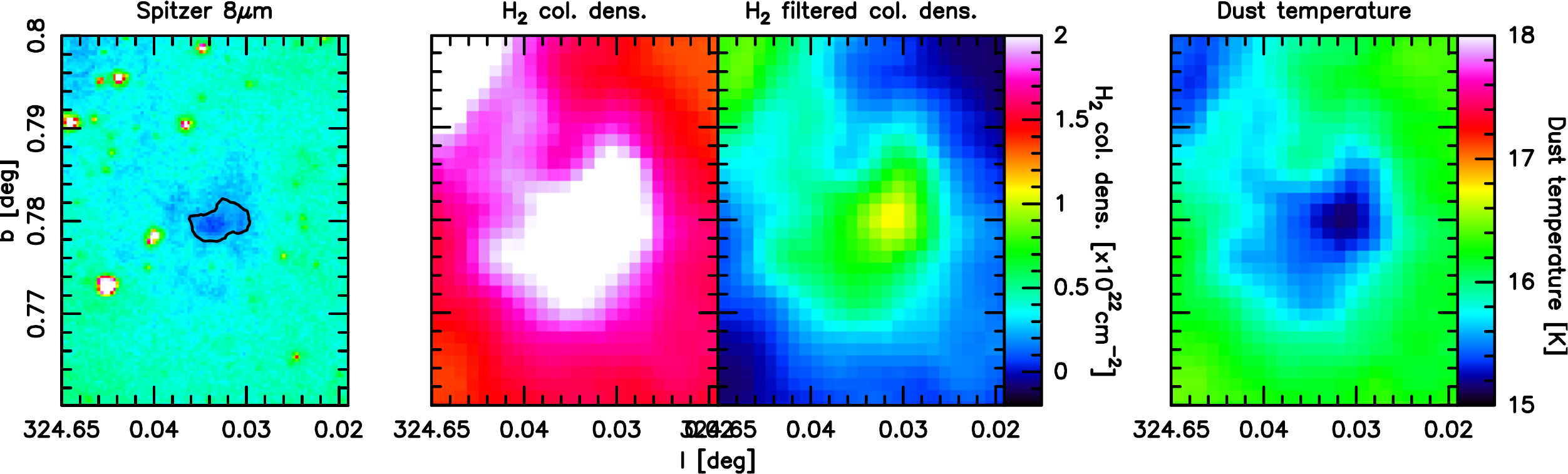}
\caption{Images of SDC324.633+0.779. From left to right: \spitzer\ 8\microns\ image with the $\tau_{8\mu m }=0.35$  opacity contour showing the boundary of the SDC as defined in PF09; \herschel\ H$_2$ column density map with background;  \herschel\  background filtered H$_2$ column density map;  \herschel\  dust temperature map.
 }
\label{irdc}
\end{figure*}

\section{Introduction}

Only in recent years have technological breakthroughs made
far-infrared/sub-millimetre Galactic plane surveys at sub-arcminute
resolution possible \citep{schuller2009, molinari2010,aguire2011}. These surveys have, for the first time, the sensitivity and resolution to probe the individual dust clumps in which stars form, providing a high resolution view of the star formation process on a Galactic scale. In particular, a complete understanding of the origin and distribution of stellar masses is only possible with large surveys of clumps which sample the full range of different physical properties and in which the initial conditions for star formation are still imprinted. In this context, performing a Galactic plane survey of infrared dark clouds (IRDCs)  is essential.

IRDCs were first observed in 1996 by \citeauthor{perault1996} using ISOCAM at 15$\mu$m as absorption features against the infrared background of the Galactic plane. Since then, follow-up observations have shown that these sources are cold, dense molecular clouds and potential mass reservoirs for future generations of stars \citep[e.g.][]{teyssier2002, simon2006a,   rathborne2006, ragan2009}. Their darkness in the mid-infrared domain ensures that these sources represent early stages of dense cloud evolution, and as such, IRDCs might contain the initial conditions of star formation.  \citet[paper I, hereafter PF09]{peretto2009} constructed a catalogue of over 11,000 \spitzer\ dark clouds (SDCs) using the 8\microns\ GLIMPSE Galactic plane survey \citep{churchwell2009} covering the $10\degree<|l|<65\degree$, $|b|<1\degree$ region (see Fig.~\ref{irdc} for an example of a SDC). In PF09,  4\arcsec\ angular resolution column density maps were constructed from the 8\microns\ extinction for all SDCs. This database has been used since for follow-up observations of specific clouds \citep{peretto2010b,peretto2013,peretto2014}, but also to tackle Galactic-scale problematics such as the mass distribution of IRDCs and their sub-structures \citep{peretto2010a}, the existence of column density thresholds for the formation of massive stars \citep{kauffmann2010}, or the characterisation of massive dense clumps \citep{traficante2015}. This type of global study is the main reason behind building the PF09 catalogue in the first place. However, the PF09 catalogue is contaminated by spurious clouds since any significant dip in the 8\microns\ emission of the Galactic plane on angular scales lower than $\sim5\arcmin$ is considered to be the result of the extinction by a cloud while it could simply be due to the intrinsic variation of the Galactic plane emission. Disentangling between real and spurious SDCs is therefore crucial for any Galactic-scale study that makes use of the PF09 catalogue.

\citet{wilcock2012}  looked at the $300\degr < l < 330\degr$ region of the Galactic plane using Herschel Hi-GAL data and visually estimated that only 38\% of the dark clouds from the PF09 catalogue were bright at 250\microns.  Taken at face value, this suggests that most of the catalogued SDCs are artefacts, only a minority of them are real, casting doubts on any global-scale study that is based on the entire PF09 catalogue. However, a more rigorous approach to cloud identification is to systematically assess which SDCs are associated with peaks in the H$_2$ column density determined from Hi-GAL data. This is the main objective of the present study.

In this paper, we present an analysis of the \herschel\ counterparts of all SDCs from the PF09 catalogue based on H$_2$ column density images. Section 2 presents the observations.  Section 3 describes how \emph{Herschel} column density maps are constructed. Section 4 explains the identification scheme and its limitations.  Finally, we discuss our results in Section 5.\\

\section{ \emph{Herschel} data}

In order to confirm the nature of the SDCs, we use far-infrared dust emission
data taken with the \emph{Herschel} space observatory \citep{pilbratt2010}. The two onboard
photometry instruments, PACS \citep{poglitsch2010} and SPIRE \citep{griffin2010}, allow the
simultaneous observation of the dust emission at 5 wavelengths
in the range 70-500$\mu$m. The Hi-GAL open time key project \citep{molinari2010}
has observed the entire Galactic plane for a Galactic latitude $|b|<1\degr$  at wavelengths of 70, 160, 250, 350 and 500$\mu$m, 
offering a unique opportunity to study the dust emission
properties of SDCs.

The Hi-GAL data were reduced, as described in \citet{traficante2011}, using
HIPE \citep{ott2010} for calibration and deglitching (SPIRE only), routines
specially developed for Hi-GAL data reduction (drift removal, deglitching),
and the ROMAGAL map making algorithm. Post-processing on the maps has been applied to help with image artefact removal \citep{piazzo2015}. In this paper, we make use of only the
PACS 160\microns\ and SPIRE 250\microns\ data, with a nominal angular
resolution of $\theta_{160}=12$\arcsec\ and $\theta_{250}=18$\arcsec, respectively.

In addition, zero-flux levels for every Hi-GAL field have been recovered by correlating \herschel\ data with Planck and IRAS data \citep{bernard2010}.

\section{Column density from \emph{Herschel}}
\label{sec_coldens}

The main goal of this study is to disentangle real from spurious SDCs in the PF09 catalogue. We 
believe that \herschel\ column density maps are probably the most suited data to do so. In this section , we discuss the construction and reliability of our  \herschel\ column density maps.

\subsection{Building 18\arcsec\ Herschel column density maps from the 160\microns/250\microns\ colour}

A difference in angular resolution is a major issue when cross-correlating
two samples of sources. For this reason, it is essential here that we
construct the highest possible angular resolution column density maps using the
\emph{Herschel} data. The typical way to construct \emph{Herschel} column
density maps is to perform pixel-by-pixel SED fitting using \emph{Herschel}
data at 4 or 5 wavelengths
\citep[e.g.][]{peretto2010b,battersby2011}. While this is the most
reliable way of constructing such maps, it requires that the data is smoothed to the resolution at the longest
wavelength, i.e. $\sim36\arcsec\ $at 500\microns.
For the purpose of confirming whether a SDC corresponds to a column density
peak,  a simpler, faster, analysis can be used, one which also
produces higher angular resolution column density maps .

\begin{figure}[t]
\centering
\includegraphics[width=7cm]{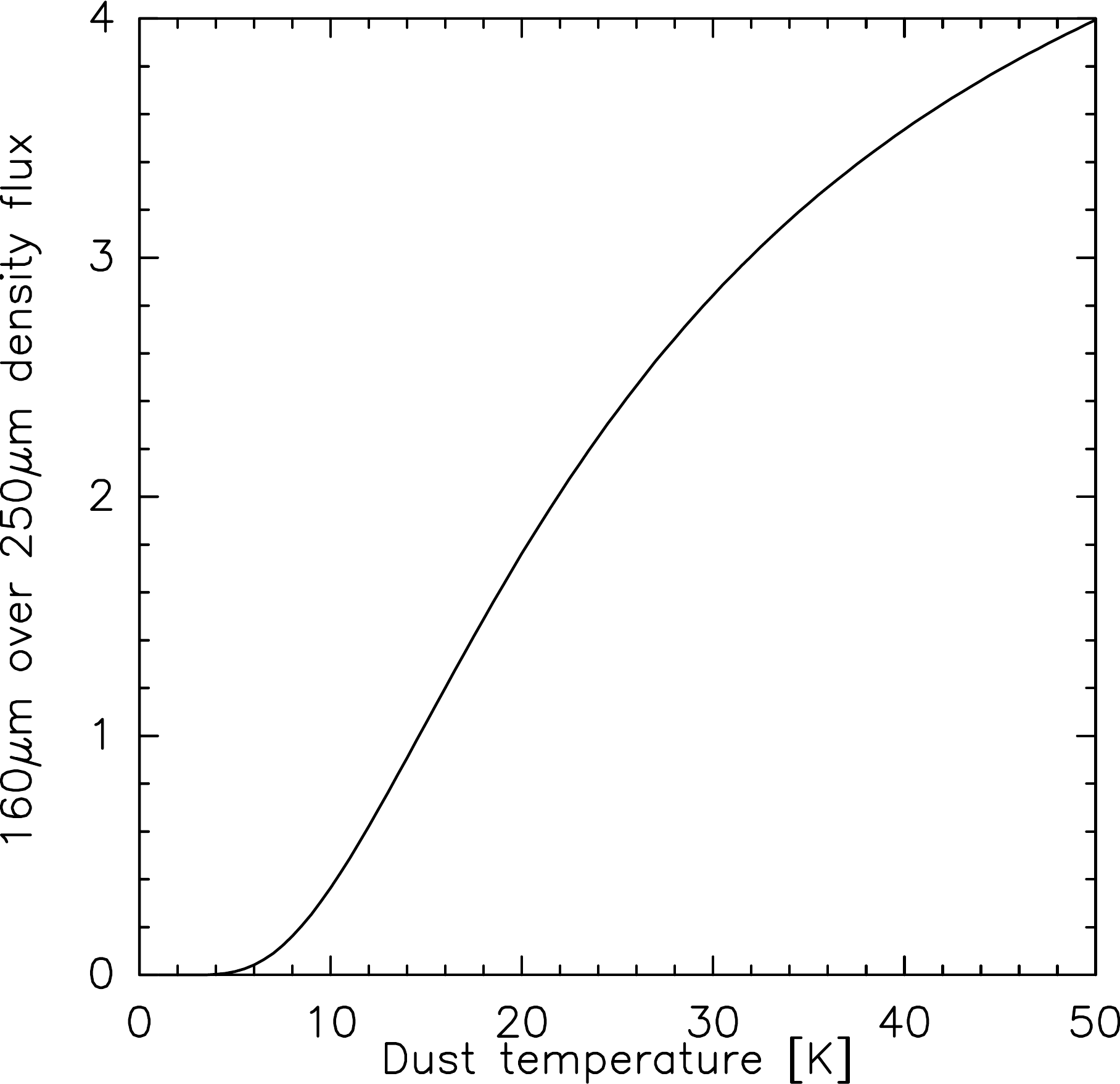}
\caption{Plot showing the variation of $R_{160/250}$, the ratio
  between 160$\mu$m to 250$\mu$m flux density as a function of dust
  temperature (see Eq.~\ref{eq_ratio}).  A value of 2 was adopted for the spectral index of the dust
  opacity law (i.e. $\beta$). }
\label{pic_temp_ratio}
\end{figure}

Here we use the ratio of the Hi-GAL 160\microns\ over 250\microns\ images as a
temperature tracer, and use the derived temperature to estimate the column density
from the 250\microns\ data. The 160\microns\ to 250\microns\ flux ratio,
$R_{160/250}$, can be written as:
\begin{equation}
\label{eq_ratio}
R_{160/250} = \frac{S_{160}}{S_{250}}=\frac{B_{\nu_{160}}(T_d)}{B_{\nu_{250}}(T_d)}\left(\frac{250}{160}\right)^{\beta}
\end{equation}
where $S_{\lambda}$ is the flux density at the wavelength $\lambda$,
$B_{\nu}$ is the Planck function, $T_d$ is the dust temperature, and
$\beta$ is the spectral index of the specific dust opacity law, and set to 2
\citep{hildebrand1983}.  As shown in Fig.~\ref{pic_temp_ratio}, $R_{160/250}$
is a monotonic function of the dust temperature, and therefore can be used to
estimate the dust temperature. In the 10 - 20~K temperature range, typical of
IRDCs \citep{peretto2010b}, this ratio varies by a factor of $\sim5$.  In practice, because the 160\microns\ and 250\microns\  images have originally different pixels sizes and projection centres, we regridded the 160\microns\ images to match the 250\microns\ image astrometry. We then convolved the 160\microns\ images to the 250\microns\ image resolution using a Gaussian kernel of FWHM $\theta_{ker}=\sqrt{\theta_{250}^2-\theta_{160}^2}=13.4$\arcsec\ . We used the resulting convolved 160\microns\ image with the original 250\microns\ image to compute the $R_{160/250}$ ratio maps. We
then converted the \higal $R_{160/250}$ maps into a temperature
map.  Note that the signal to noise ratio in Hi-GAL maps is very high, with a minimum value of 10 at 160 and 250\microns\ for the faintest regions of the Galactic plane covered by \herschel\ (Molinari et al., submitted). This means that for the vast majority of the SDCs studies here, the uncertainty on $R_{160/250}$ is only a few percents, which translates into a temperature uncertainty of a few tenths of a Kelvin. 
To calculate the column density map, we then combine this temperature
map with the Hi-GAL 250\microns\ image, to derive the column density through
the equation:
\begin{equation}
\label{eq_coldens}
 N_{H_2} = S_{250} / [ B_{\nu_{250}} (T_d)\kappa_{250} \mu m_H] 
\end{equation}
where $\rm \kappa_{250}=0.12cm^2 g^{-1}$  \citep{ossenkopf1994} is the specific
dust opacity at 250\microns\  (that already includes a dust to gas mass ratio of 1\%), $\mu=2.33$ is average molecular weight, and $m_H$
the atomic mass of hydrogen. 
The simplicity of this method allows the rapid
construction of relatively high (i.e. 18\arcsec) angular resolution column density maps for all the SDCs (see Fig.~\ref{irdc}).

\begin{figure}[ht]
\centering
\includegraphics[width=9cm]{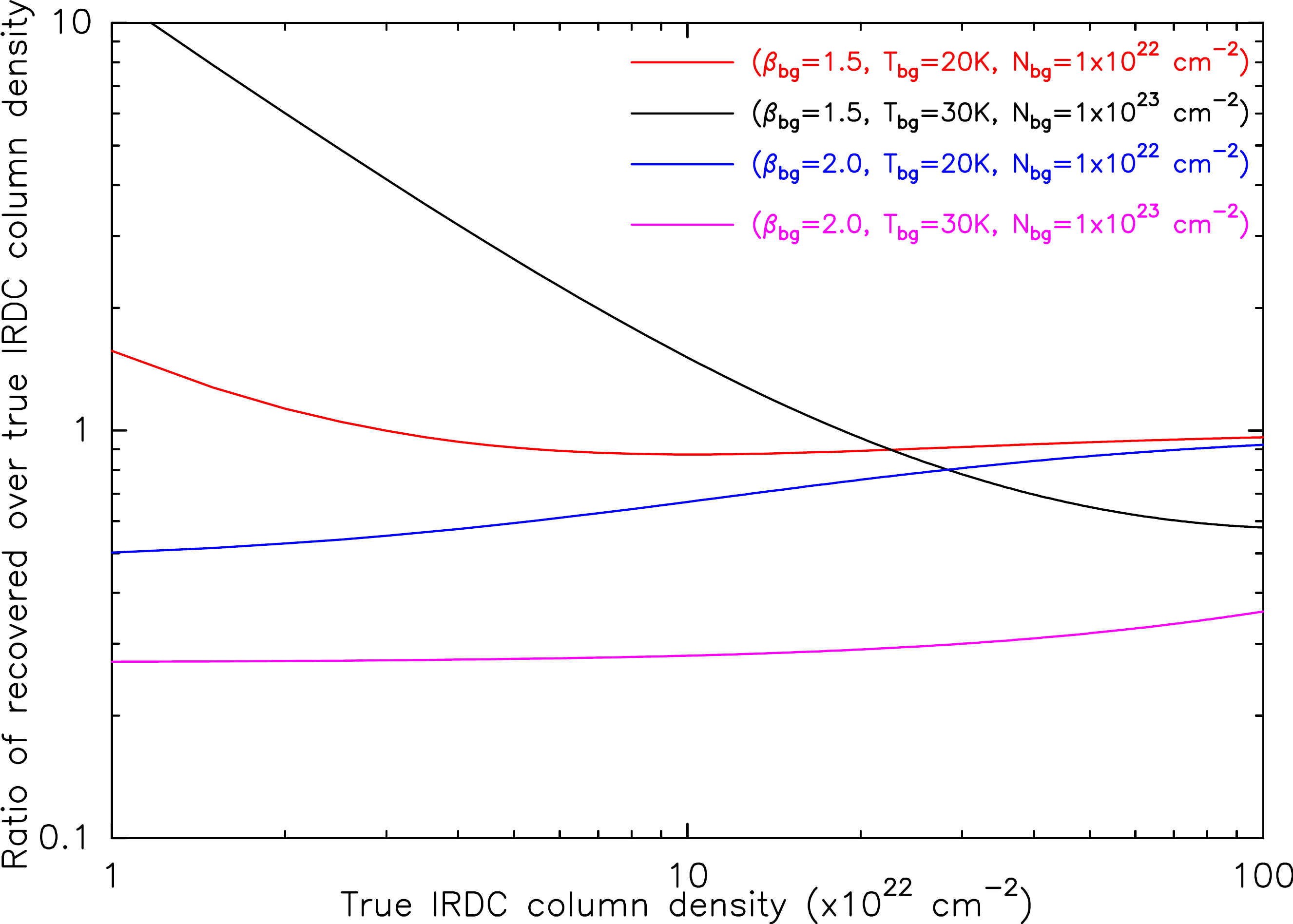}
\caption{Plot showing the uncertainty linked to our column density construction method.  On this plot we show the input IRDC column density versus the ratio of the retrieved column density versus true IRDC column density. The different colour correspond to different background properties. 
}
\label{simflux}
\end{figure}

Note that the column density we measure toward an IRDC is integrated along the
line of sight and is the sum of the dust column density from the IRDC and the
warmer column density from the background. These two components have potentially different dust properties (temperature and specific opacities), and column densities. In some places, the background
column density can be larger than the column density of the cloud itself. It is possible to reconstruct the background first and remove its contribution to the observed fluxed towards the IRDCs \citep{peretto2010b,battersby2011}. However, this is a difficult task for
such a large sample of objects. For this
reason, we decided to use a more practical method that filters out large scale structures in the column 
density map constructed as described above. For this purpose we used  a 10\arcmin\ wide median filter on the {\it Herschel} column density maps to create a background image, and subtracted this median
component from the original column density image to create a background-subtracted column density map  (see Fig.~\ref{irdc}).  These are the maps that we used for the remainder of the analysis. The width of the filter was chosen so that it is of similar size as the largest SDCs of the catalogue.

\subsection{The impact of background and SDC dust emission mixing on retrieved cloud properties}

Estimating the column density without separating the background and IRDC contributions to the flux densities  could lead to errors on the retrieved column density and temperature of IRDCs. In order to quantify this error we modelled the emission of a background and IRDC components as modified-blackbodies at different temperatures and column densities, and added their respective flux densities at both 160\microns\ and 250\microns. We then used the same procedure as outlined in the previous section to estimate the temperature and column density of the combined IRDC/background components. We finally removed the original background column density  from the combined column density to retrieve the IRDC column density.  We varied both the properties (column density and dust emissivity index) of the background and the column density of the IRDC itself (with a constant temperature of 12K).  In Fig.~\ref{simflux} we show the ratio of the recovered IRDC column density using this technique over the input IRDC column density for different background/IRDC properties. One can see that the errors on the IRDC column densities can be quite high (up to a factor of 10 for low column density IRDCs) for warm and high  column density background. However, for more typical background properties, the errors are within a factor of 2. In all cases, the temperature of the IRDC is overestimated, by only a few tenths of a Kelvin in the best cases, and up to 10~K in the most difficult cases (low-column density IRDCs against high-column density and warm background).

\begin{figure*}[!ht!]
\centering
\includegraphics[width=18.5cm]{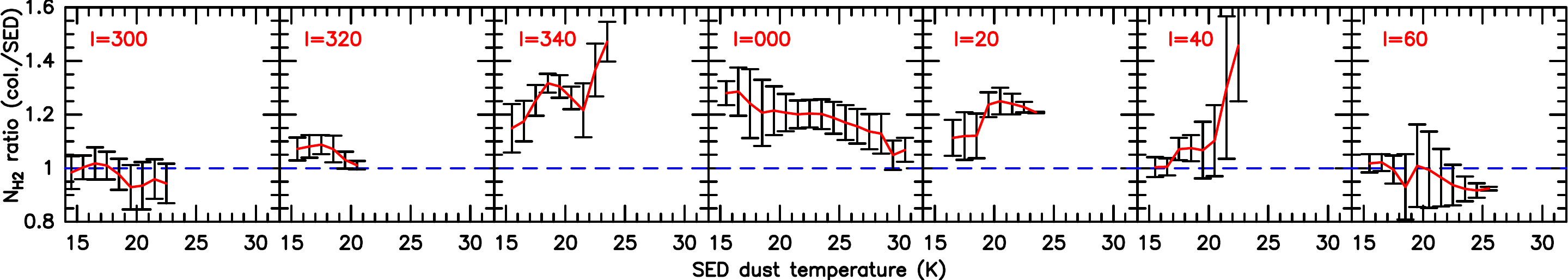}
\caption{Figure showing the ratio of the colour-based over SED column densities as a function of SED dust temperature for entire Hi-GAL tiles. The red solid line shows the average ratio over the corresponding tile, while the error bars represent one $\sigma$ deviations. The blue dashed line indicates a ratio of 1.
} 
\label{ratiocd}
\end{figure*}

\subsection{The relative uncertainty of colour versus SED column density maps}

To test further our method for calculating the IRDC column densities, we compared our 160\microns/250\microns\ colour column densities to the more standard 4 points [160, 250, 350, 500\microns] SED fitting technique. We computed the column densities following the two methods for entire Hi-GAL tiles at six different locations in the Galaxy and made a pixel-by-pixel ratio of the resulting column densities after convolving our colour column density map to the same 36\arcsec\ resolution of the SED column density map. Figure \ref{ratiocd} shows how this ratio varies as a function of dust temperature for all tiles. In this plot, IRDCs correspond to the points at lowest temperatures. We can see that the agreement between the two techniques remains within $\sim 30\%$ in most cases. The agreement is even better for typical SDC temperatures ($<20$~K), and improves when moving away from the Galactic centre. The different trends observed can probably be explained with changes in dust properties, but a full investigation on this is beyond the scope of this paper.

Overall, the method we use to calculate the column density is probably accurate within a factor of 2 for most IRDCs, uncertainties being dominated by the background/IRDC component separation. These uncertainties do not include systematic uncertainties on the dust emissivity which can account for an extra factor of 2.

\begin{figure}[t]
\centering
\includegraphics[width=7cm]{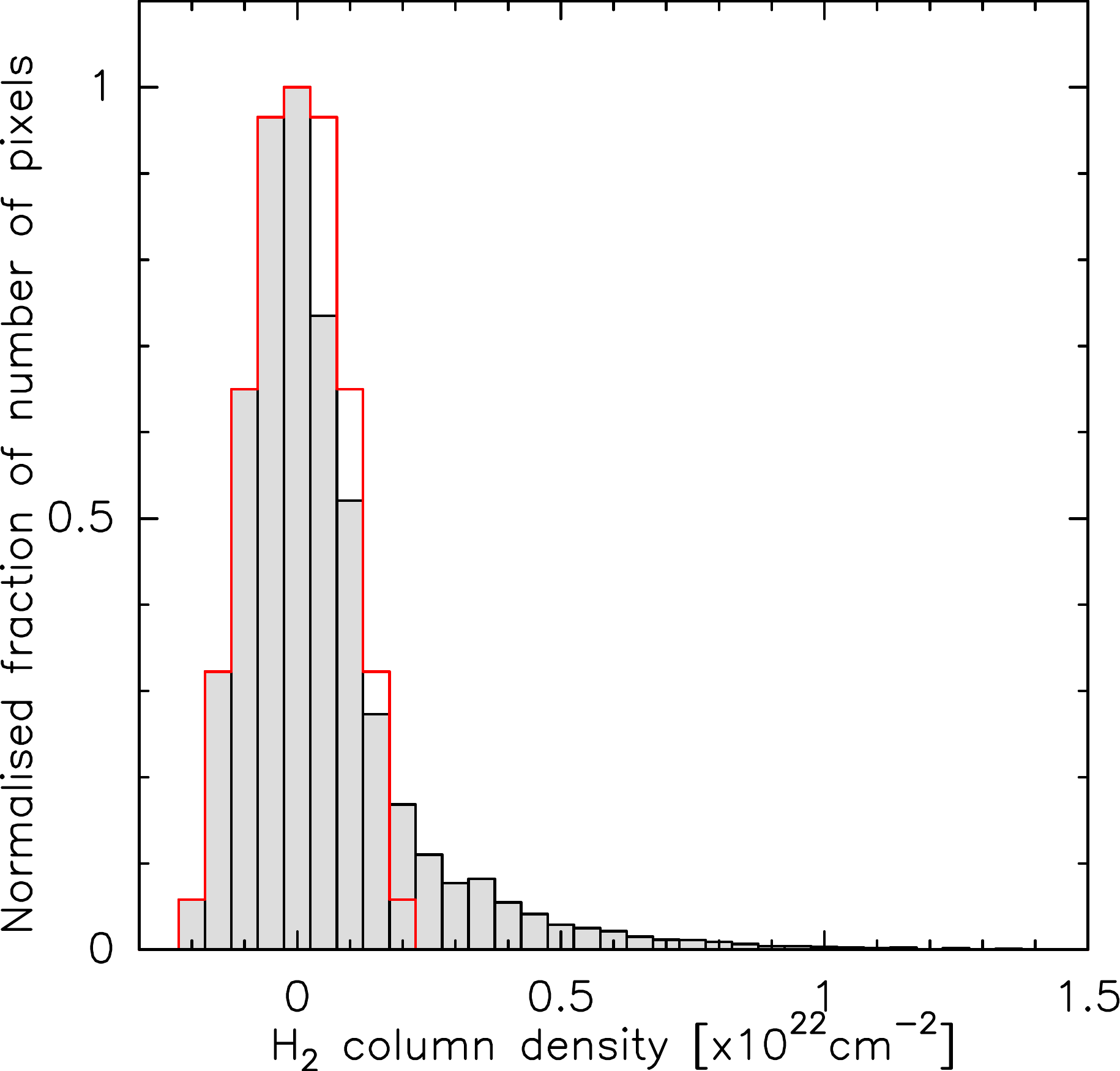}
\caption{Histogram of background subtracted column density pixels (grey histogram) within a 10\arcmin\ box centred on the central pixel displayed in Fig.~\ref{irdc}. The column density noise estimated in each pixel ($\sigma_j$) is corresponds to the dispersion of the red histogram, obtained by mirroring the negative part of the original histogram about its peak. 
} \label{histonoise}
\end{figure}

\section{SDC detection in \emph{Herschel} column density maps}
\label{sec:higal_detect}

\subsection{Detection criteria}

In order to identify which SDCs are detectable in the \emph{Herschel} column
density maps, we first computed a map of column density noise. This is
constructed by computing the histogram of the background-subtracted column
density pixels in a 10\arcmin\ box centred on each pixel. Then, in a similar
manner as \citet{battersby2011} we compute the dispersion $\sigma_j$ of the
background-subtracted Hi-GAL images by mirroring the negative values about the
histogram peak, and measure the dispersion on the resulting histogram for pixel $j$ (see Fig.~\ref{histonoise}). This dispersion is representative of the column density fluctuations of the
background on scales lower than 10\arcmin.

In order to decide whether an IRDC is real  we defined three criteria. The first one, $c_1$, is the difference
between the average \herschel\ column density within the $\tau_{8\mu m}$ boundary (as defined in \citet{peretto2009}, see Fig.~\ref{irdc}) of the IRDC, $\overline{N_{H_2}^{in}}$, and the average \herschel\ column density immediately outside this boundary, $\overline{N_{H_2}^{ out}}$. By immediately outside we mean within the rectangular cutouts that have been defined in \cite{peretto2009} to extract every SDC. The dimensions of these cutouts are twice the size of the SDC in both x and y directions (i.e. the image axes). If  the IRDC is real, then we expect:

\begin{equation}
 c_1=\left(\overline{N_{H_2}^{in}} -\overline{N_{H_2}^{out}}\right)>0
 \end{equation}
  The second parameter, $c_2$, is defined as:
  \begin{equation}
  c_2=\overline{N_{H_2}^{in}}/\overline{\sigma_{in}} \ge 3
  \end{equation}
   where $\overline{\sigma_{in}}$ is the column density dispersion estimated on the scale of the cloud defined as:

 \begin{equation}
 \begin{array}{lr}
  \overline{\sigma_{in}}=<\sigma_j>=\frac{\sum\limits_{j=1}^{n_{pix}}\sigma_j}{n_{pix}}   & \text{if }  R_{eq} < 9\arcsec\\
 &\\
 \overline{\sigma_{in}}=\frac{<\sigma_j>}{\sqrt{n_{beam}}}  = \frac{\theta_{beam}}{2\sqrt{\ln(2)}R_{eq}}\frac{\sum\limits_{j=1}^{n_{pix}}\sigma_j}{n_{pix}}  & \text{if } R_{eq}\ge 9\arcsec 
 \end{array}
 \end{equation}
 where, $n_{pix}$ is the number of pixels within the boundary of the IRDC, $n_{beam}$ is the number of \herschel\ beams within the IRDC boundaries,  $\theta_{beam}=18\arcsec$ is the resolution of the \herschel\ column density maps,  and $R_{eq}$ is the equivalent angular radius of the IRDC (as defined in \citet{peretto2009}).

  The third criterion, $c_3$,  is defined as:
 
 \begin{equation}
  c_3=\left(\overline{N_{H_2}^{in}} -\overline{N_{H_2}^{out}}\right)/\overline{\sigma_{in}}\ge3
  \end{equation}

   This last criterion is more selective than $c_2$ and, as a result of eye-investigation, we consider that a significant number of real IRDCs would be missed by using it, picking up very high signal to noise ratio clouds (an example of a SDC meeting $c_1$ and $c_2$ criteria but failing $c_3$ is shown in the Appendix A - SDC15.422-0.098). The detection results presented in this paper are based on $c_1$ and $c_2$ only.
    
   For each SDC, Table \ref{summary} gives: the name, angular radius, $\overline{N_{H_2}^{in}}$, $\overline{N_{H_2}^{out}}$, $ \overline{\sigma_{in}}$,  the peak H$_2$ column density estimated with \herschel\  within the SDC boundary $N_{H_2}^{pk}$, $c_1$ value, $c_2$ value, $c_3$ value, and  the last column indicates if the clouds satisfy the $c_1$ and $c_2$ criteria.

 \begin{figure}[t]
\centering
\includegraphics[width=9cm]{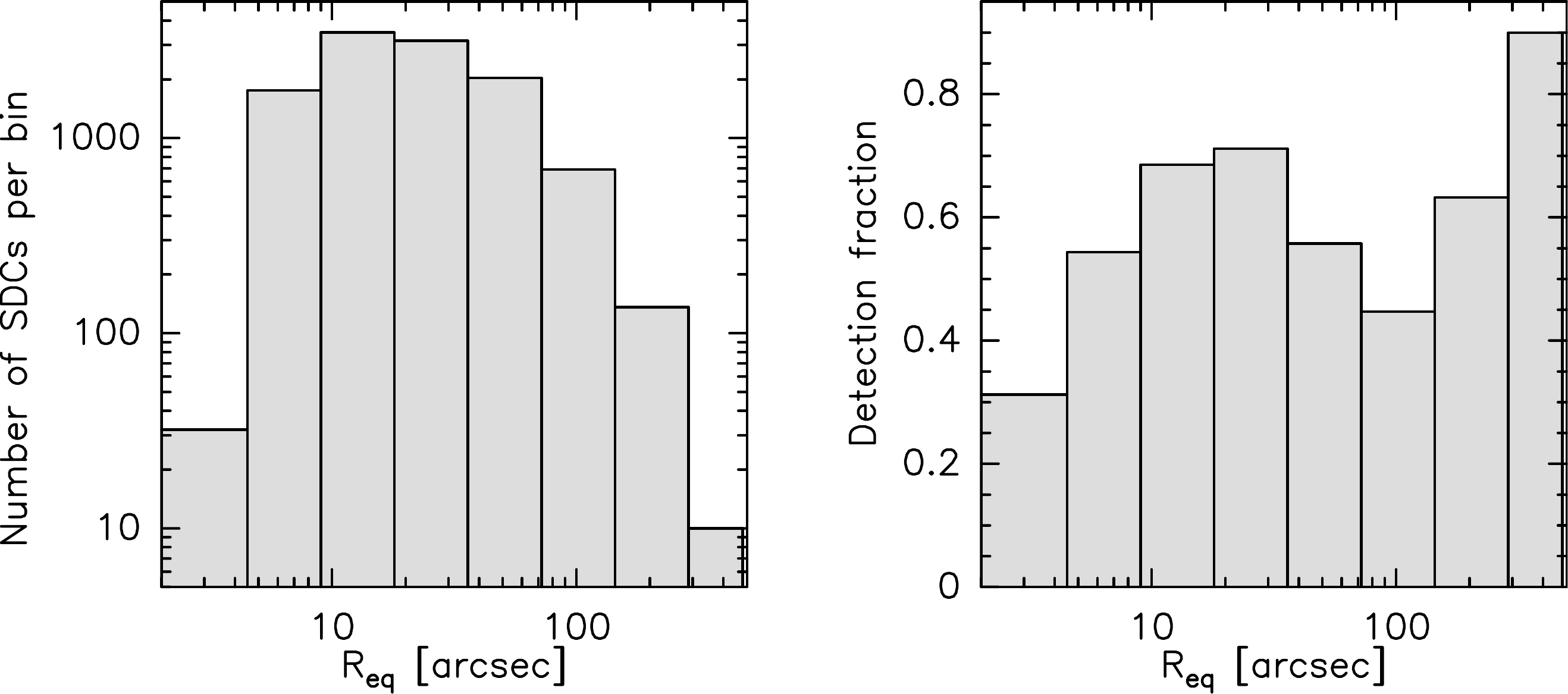}
\caption{Number of SDCs (left) and  automated detection fraction (right) as a function of SDC angular radius.
} 
\label{detect}
\end{figure}

\begin{table*}
\caption{SDC \herschel\ counterpart properties. The full table is available at the CDS.}             
\label{summary}      
\centering                          
\begin{tabular}{c c c c c c c c c c}        
\hline
\hline                 
Name   & $R_{eq}$  & $\overline{N_{H_2}^{in}}$ & $\overline{N_{H_2}^{out}}$  & $\overline{\sigma_{in}}$ & $N_{H_2}^{pk}$  & $c_1$ &  $c_2$ &  $c_3$ & Detected?  \\  
 &   \arcsec &  $\times10^{22} cm^{-2}$  & $\times10^{22} cm^{-2}$ & $\times10^{22} cm^{-2}$ & $\times10^{22} cm^{-2}$& &  & & \\  
 \hline   
 SDC10.014-0.818 &    65.4 &     -0.10 &    -0.03 &    0.02 & 0.19  &-0.07 & -4.3 & -3.0 & n \\  
SDC10.031-0.355 &  14.7   &  5.09     &  	2.96 	  &  0.25   & 5.95  &2.13 & 20.0 & 8.4 & y \\
SDC10.043-0.425 &  62.2  &  1.19     & 0.71        & 0.06    & 4.69 &0.48 & 21.3 & 8.6 & y\\
SDC10.055-0.355 &  20.1  &  1.43     &  0.71       & 0.21    & 2.30  &0.72 & 6.8   & 3.4 & y \\
SDC10.067-0.406 &   6.9   & 4.70      & 3.26        &  0.37   & 5.82 &1.43  & 12.8  & 3.9 & y \\
SDC10.069-0.400 &  4.2   &  2.83     & 2.59        &    0.38  & 3.65 &0.24  & 7.4   & 0.6  & y \\
 SDC10.082-0.414 & 25.8  & 2.20    & 1.21       & 0.15     &   3.56 &0.98   & 14.4 & 6.4 & y \\
 SDC10.086-0.438 & 38.0    & 1.06   & 0.60        & 0.09     & 2.24  &0.46  & 12.2  & 5.3 & y \\
 SDC10.094-0.415  & 3.8    &  1.49    & 1.48       & 0.35     & 1.62 &0.01  & 4.2    & 0.0 & y \\
 SDC10.111-0.431  & 12.0   &  0.89     & 1.11     & 0.28    & 1.32  &-0.22  & 3.2  & -0.8  & n \\
    \hline                                   
\end{tabular}
\end{table*}

The PF09 catalogue contains 6 SDCs that are not covered by Hi-GAL, and one which is located in a saturated portion of the 250\microns\ \herschel\ data, leaving a total of 11,289 SDCs for analysis.  
Figure~\ref{detect} shows the histogram of SDC angular radius\footnote{The radii used here and quoted in Table~\ref{summary} differ slightly from the ones in PF09. As a result of cloud reprojection a mistake had  been made on the size of the pixel of the \spitzer\ images, which reflected in an over-estimate of the SDC radii up to 30\%.}, along with the histogram of the fraction of clouds satisfying criteria $c_1$ and $c_2$ per SDC size bin. For the remainder of this paper, we will refer to this fraction as {\it automated detection fraction}.

Using criteria $c_1$ and $c_2$, 63.2\% (7,136) of the 11,289 SDCs of the PF09 catalogue are detected with \herschel. One can see on Fig.~\ref{detect} that this detection fraction varies as a function of radius, with an increase up to an angular radius of 30\arcsec , then a decrease up to $R_{eq}\simeq 100$\arcsec, and a final increase at larger radius. This detection curve is affected by a number of elements which impact its interpretation. In order to get a better insight into Fig.~\ref{detect} we decided to simulate the SDC detection process.

\begin{figure}[ht]
\centering
\includegraphics[width=6.5cm]{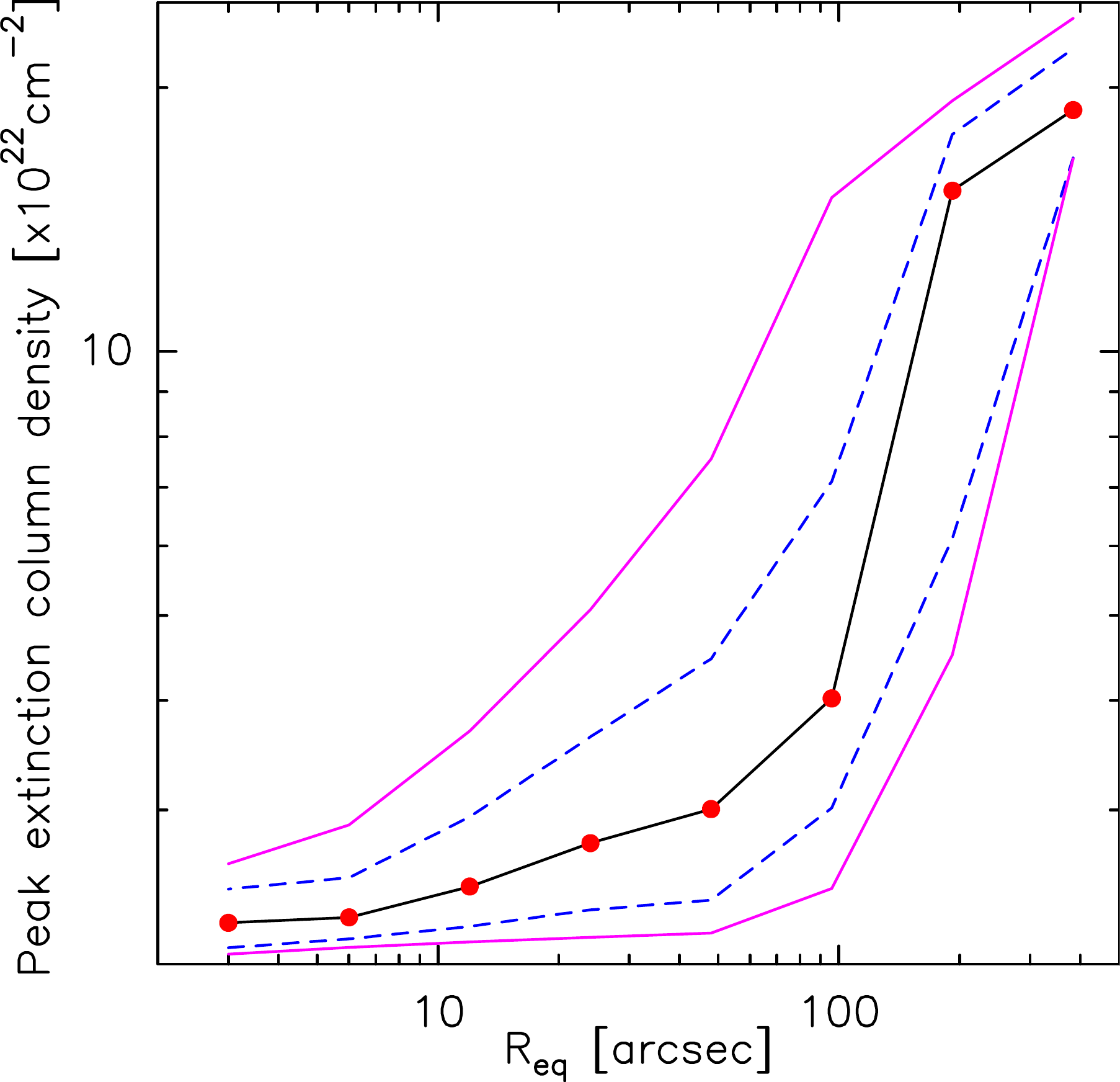}
\caption{Angular radius binned the same way as in Fig.~\ref{detect} versus the median (red symbols and black solid line) peak extinction column densities of all SDCs from the PF09 catalogue. The purple solid lines  and blue dashed lines are the 10/90 percentiles and 25/75 percentiles, respectively.
} 
\label{req_medcd}
\end{figure}

\subsection{Detection simulations}

Because of the resolution difference between \emph{Herschel} at 250$\mu$m
(18\arcsec) and \spitzer\ at 8$\mu$m ($\sim2\arcsec$), along with the strong
fluctuations of the Galactic column density background, some of the smaller
real SDCs may be missed by our identification scheme while others might be wrongly classified as real.  
To evaluate the impact of background variations
on our detection scheme, and therefore have a better estimate of the fraction of spurious clouds, we performed
simulations of our cloud detection method.

Instead of taking idealised cloud models (such as Bonnor-Ebert spheres for instance) we decided to  use SDCs themselves, as we believe they provide  a more representative view of detection outcome. It is clear that large clouds with large column density peaks will be more easily detected than small and low column density ones. In order to get a sample of SDCs that is representative of the full SDC population, we first computed the angular size versus peak column density (from extinction) for the SDCs from the PF09 catalogue. This is shown in Fig.~\ref{req_medcd}. We can see that the peak column density is smoothly increasing up to $R_{eq}\simeq50\arcsec$ and then increases rather sharply. The last two, and potentially even last three, points of this plot are heavily contaminated by spurious clouds though (cf below). Given this contamination it seems likely that for real clouds the trend observed below $R_{eq}=50\arcsec$ continues to larger sizes.  In any case, as we show below, all real clouds beyond an angular radius of 60\arcsec\ should be detected, no matter what. So for the modelling we selected a sample of 6 SDCs whose sizes and column densities follow the median curve (black solid line) of Fig.~\ref{req_medcd}, up to $R_{eq}=60\arcsec$.

 \begin{table*}
\caption{SDC visual inspection summary}             
\label{visual}      
\centering                          
\begin{tabular}{c c c c c c c c c c}        
\hline\hline                 
Size bin   & Nb of SDCs &$N_{real}^{visu}$ & $N_{spur}^{visu}$ &$N_{real}^{auto}$ & $N_{spur}^{auto}$ & $n_{real}$ & $n_{spur}$ &$f_{real}$ & $f_{spur}$ \\  
 &  &  \% &  \%  &  \% & \% & \% &  \%  &  \% &  \%\\  
 \hline                        
  8\arcsec-16\arcsec    &  50  & 88&12 & 66 & 34 &  22& 0 &  75& 0 \\
  16\arcsec-32\arcsec    &  50  & 82 &18 & 66 & 34 & 10& 0 & 88 & 0\\
 32\arcsec-64\arcsec     &   50 & 60 & 40 & 60 & 40 & 4 & 4 & 93  & 10  \\
 64\arcsec-128\arcsec    &  50  &  40  & 60 &52  &48 & 0& 12 & 100 & 20 \\
 128\arcsec-256\arcsec   &  136  & 48 & 52 &64  & 36 & 0& 16 & 100 & 31\\
 256\arcsec-512\arcsec  &  10  &  40 & 60 & 90 & 10 & 0 & 50 & 100 & 83 \\
 \hline                                   
\end{tabular}
\end{table*}

 Using the PF09 \spitzer\ H$_2$ column density maps, assuming a uniform temperature of either 12K or 15K (which is representative of the dust temperature of such clouds, Peretto et al. 2010) and the same dust opacity law and molecular weight as in Sec.~2, we inverted Eq.~(2) to simulate the appearance of these six clouds  at  both 160 and 250\microns. We convolved these images to the \emph{Herschel} resolutions, and  placed all 6 SDCs at 100 different locations within one of the $\sim4$ square degree Hi-GAL tiles of the corresponding wavelength. The longitudes of each location were chosen so that to be regularly spaced, while the latitudes  were randomly drawn from a normal distribution of FWHM=1\degree\ and central position of -0.1\degree, as observed for our IRDC sample (see Fig.~11 of Peretto et al. 2009). We repeated the process for different Hi-GAL tiles between $l=10\degr$ and $l=63\degr$. Finally, we applied our entire identification scheme (i.e. construction of colour column density images and detection criteria) for every modelled cloud. Note that we also calculated $c_1$ and $c_2$ at the cloud location before adding them to the \herschel\ images  (using the same $\tau_{8\mu m}=0.35$ boundaries as for our modelled SDCs - cf Sec.~4). This allows us to estimate the probability of having a positive detection even though no SDC is present and therefore gives us a sense of the contamination of the SDC automated detection fraction by spurious features.

\begin{figure}[ht]
\centering
\includegraphics[width=9cm]{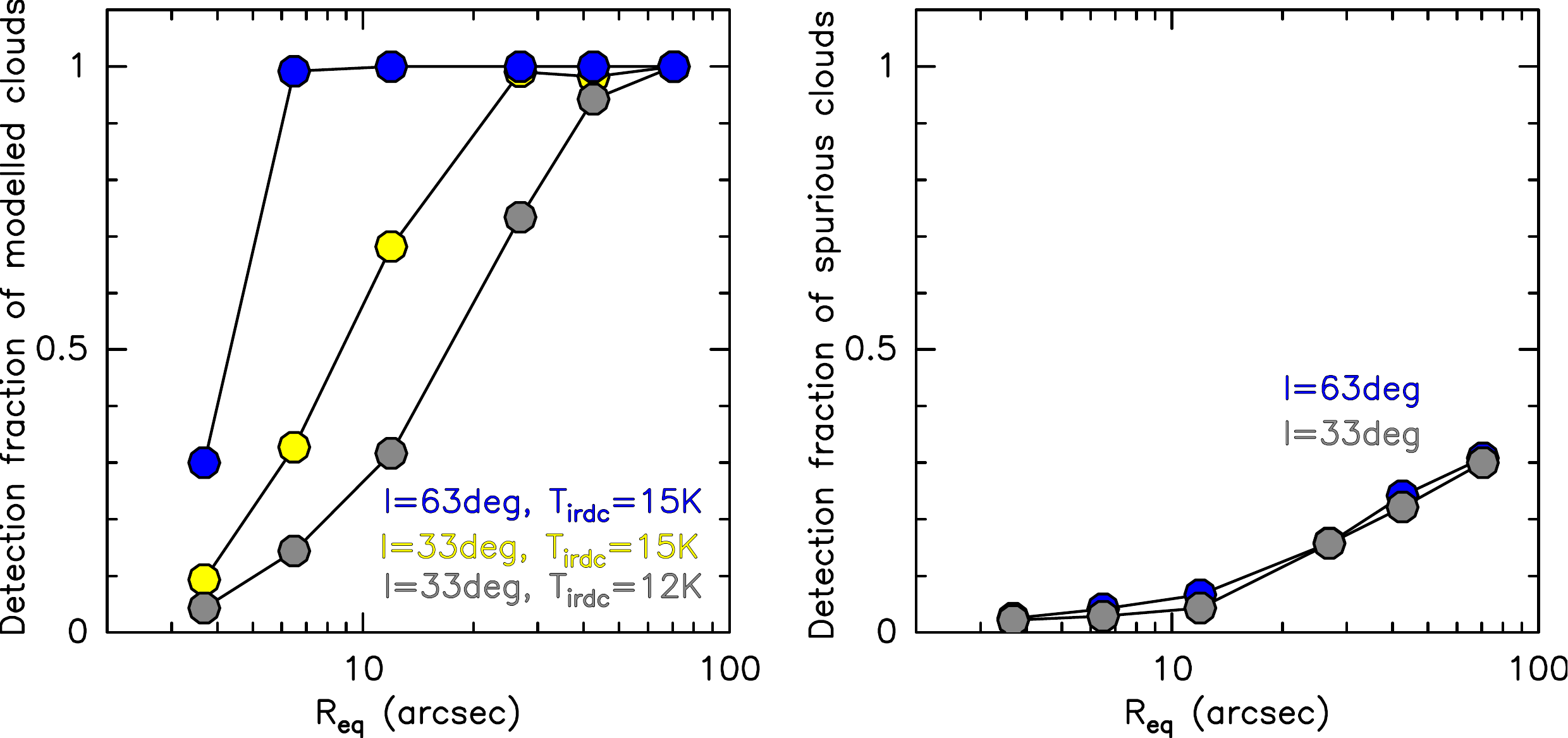}
\caption{Results of cloud detection simulations. (left): Automated detection fraction of modelled SDCs as a function of SDC size for two different locations in the Galactic plane (i.e. $l\sim33\degr$ and $l\sim63\degr$) and two different SDC dust temperature (i.e. 12K and 15K). We can see that the detection fraction strongly depends on the location (i.e. Galactic background) and cloud temperature. (right): Automated detection fraction of spurious clouds as a function of size, for two different locations. We can see see that for spurious clouds the detection behaviour is almost independent of location, and steadily increases with size. 
} 
\label{simcloud}
\end{figure}

 Figure \ref{simcloud} displays the main results of our simulations. On the left-hand-side panel, we can see that the fraction of  modelled clouds that our automated detection scheme manages to identify strongly varies as a function of cloud sizes, for all three longitudes displayed here. This fraction reaches 100\% if the cloud is larger than 60\arcsec\ independently of the cloud temperature and location in the Galactic plane. For smaller clouds the automated detection fraction depends on the cloud temperature and strength of the Galactic background (decreasing from the Galactic centre outwards). Modelled clouds with $R_{eq}\le10\arcsec$ are the most difficult to identify, with  an automated detection fraction which could be as low as 10\%. 
  
 On the right-hand-side panel of Fig.~\ref{simcloud} one can see the fraction of spurious (i.e. non-existant) clouds that manage to pass criteria $c_1$ and $c_2$ and therefore would be considered as real according to our automated detection scheme. This fraction remains below 5\% until the cloud reaches an angular radius of $\sim10\arcsec$ and then smoothly increases up to $\sim 30\%$ for clouds with $R_{eq}\simeq100\arcsec$. For the largest clouds of the PF09, the automated detection fraction of spurious clouds can reach 50\% or more. These spurious  detections are related to the probability of getting high column density peaks in a given area, i.e. the larger the area the larger the probability.  It also explains the break in the size column density plot of Fig.~\ref{req_medcd}.

These simulations demonstrate that the  automated detection fraction displayed in Fig.~\ref{detect} is not straight forward to interpret. In the following section, we will use the results of these simulations to constrain further the fraction of real SDCs in the PF09 catalogue.

\begin{figure}[t]
\centering
\includegraphics[width=9cm]{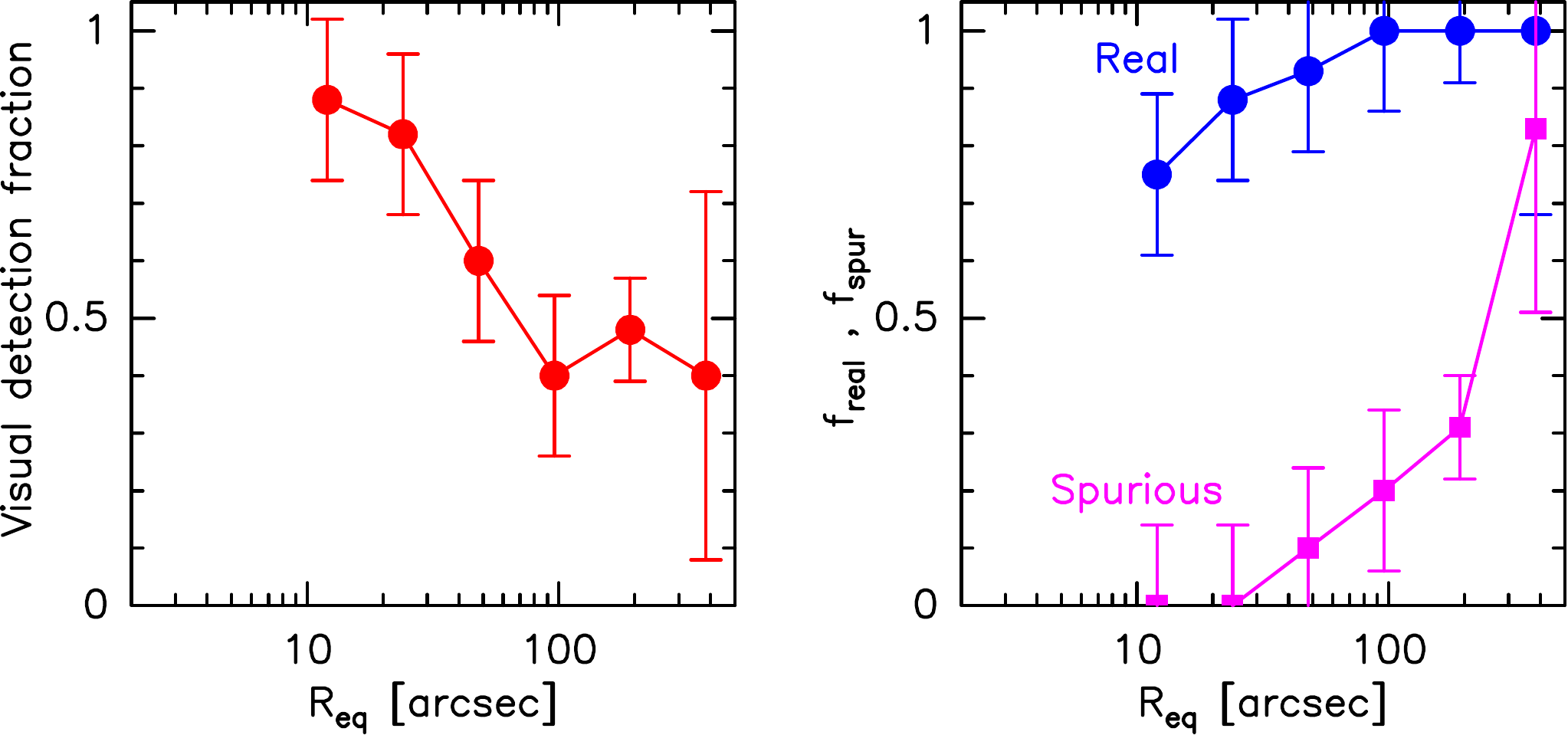}
\caption{ (Left): Fraction of real SDCs as estimated through visual inspection (red symbols and solid line). (Right): Fraction of visually-confirmed real SDCs that have been identified as real SDCs by our automated detection scheme ($f_{real}$ - blue round symbols), and fraction of visually-confirmed spurious SDCs that have been mis-classified as real SDCs by our automated detection scheme ($f_{spur}$ - purple square symbols). These two curves can directly be compared to our detection simulation results presented in Fig.~\ref{simcloud}. All error bars correspond to Poisson noise.
} 
\label{visu}
\end{figure}

\section{Discussion}

\subsection{Visual inspection}

Visually inspecting a sub-sample of SDCs is an important step towards the validation of our detection scheme and simulations. We completed this step by visually matching the morphology of the SDCs as seen in the 8\microns\ \spitzer\ images with that of their \herschel\ column density counterparts.
This can only be reliably done for rather large SDCs (i.e. of at least the size of the \herschel\ beam). We thus focused on the 6 largest size bins of Fig.~\ref{detect}.  We decided to check all 146 clouds falling in the last two bins of Fig.~\ref{detect} (10 clouds in the last bin and 136 clouds in the one before), and 50 SDCs in each of the preceding 4 size bins.  In practice, for each SDC we visually investigated, we overlaid the corresponding \herschel\ column density contours  (starting from $0.1\times10^{22}$~cm$^{-2}$ and separated by steps of  $0.5\times10^{22}$~cm$^{-2}$) on the \spitzer\ 8\microns\ image, and decided, after eye inspection, if a column density peak was convincingly matching at least a fraction of  SDC. A sample of such images is provided in Appendix A.  
The detection fraction estimated this way will be refered to as {\it visual detection fraction}.  
 
 The left-hand-side panel of Fig.~\ref{visu} displays the visual detection fraction (column $N_{real}^{visu}$ of Table 2) as red symbols. We see that at large radii, the visual detection fraction is lower than the automated detection fraction as estimated for the same cloud sub-sample (column $N_{real}^{auto}$ of Table 2).
  The fact that the fraction of spurious clouds (column $N_{spur}^{visu}$ of Table 2) increases with size is a consequence of the construction of the 8\microns\ opacity maps that are built to identify the SDCs \citep{peretto2009}. One step involves to convolve the original \spitzer\ 8\microns\ images with a 5\arcmin\ Gaussian kernel. This convolution is performed to construct the mid-infrared background image of the region. However, in places where a bright 8\microns\ region is present, this convolution artificially produces significant structures in the mid-infrared background, which translates into large spurious features in the 8\microns\ opacity maps (examples of such spurious clouds can be found in Appendix A, e.g, SDC329.368-0.437). On the other hand, at small radii, the visual detection fraction is larger. This is due to the fact that the eye can more easily identify low signal to noise sources as it recognises matching shapes in \herschel\ and \spitzer\ images. 
 
 Assuming that the visual inspection provides the true fraction of real SDCs (the $N_{real}^{visu}$ values in Table 2), we can compute  the equivalent of Fig.~\ref{simcloud} for the visually inspected sample of SDCs. For this, we need first to compute the fraction of SDCs in each size bin that have been mis-classified as spurious sources by our automated detection scheme, $n_{real}$. We also need to evaluate the fraction of SDCs that have been mis-classified as real sources by our automated detection scheme, $n_{spur}$.  With these fractions in hand one can compute $f_{real}$, the fraction of real SDCs that have been identified as real by our automated detection scheme. This is given by $f_{real}=(N_{real}^{auto}-n_{spur})/N_{real}^{visu}$, where $N_{real}^{visu}$ is the fraction of visually-confirmed SDCs (i.e. the visual detection fraction), and  $N_{real}^{auto}$ is the automated detection fraction for the same sub-sample of SDCs. This quantity is plotted as blue symbols in the right-hand-side panel of Fig.~\ref{visu}, and is directly comparable to the left-hand-side panel of Fig.~\ref{simcloud}. We can also compute $f_{spur}$, the fraction of spurious SDCs  that have been mis-classified as real SDCs by our automated detection scheme. This is given by $f_{spur}=n_{spur}/N_{spur}^{visu}$, where $N_{spur}^{visu}$ is the fraction of visually-confirmed spurious SDCs. This quantity is plotted as purple symbols in the right-hand-side panel of Fig.~\ref{visu}, and is directly comparable to the right-hand-side panel of Fig.~\ref{simcloud} (a summary of the visual inspection is given in Table 2).
 We can see that both trends (the increase of spurious detection fraction with increasing radius, and the decrease of real SDC detection fraction with decreasing radius) were predicted by our detection simulations. The amplitude of these two effects are also well reproduced. 
 
 Overall, we can reasonably say that the large majority of SDCs, with an angular radius under 60\arcsec, that have been identified as real by our automated detection scheme, are indeed real. For larger clouds, visual inspection of individual sources is required to check their nature (real versus spurious). Note as well that the extinction-based column density and therefore sizes of the largest clouds appear comparatively uncertain. If any portion of one of these clouds was clearly associated with a \herschel\ column density peak we then classified the clouds as real.

\begin{figure}[t]
\centering
\includegraphics[width=7.cm]{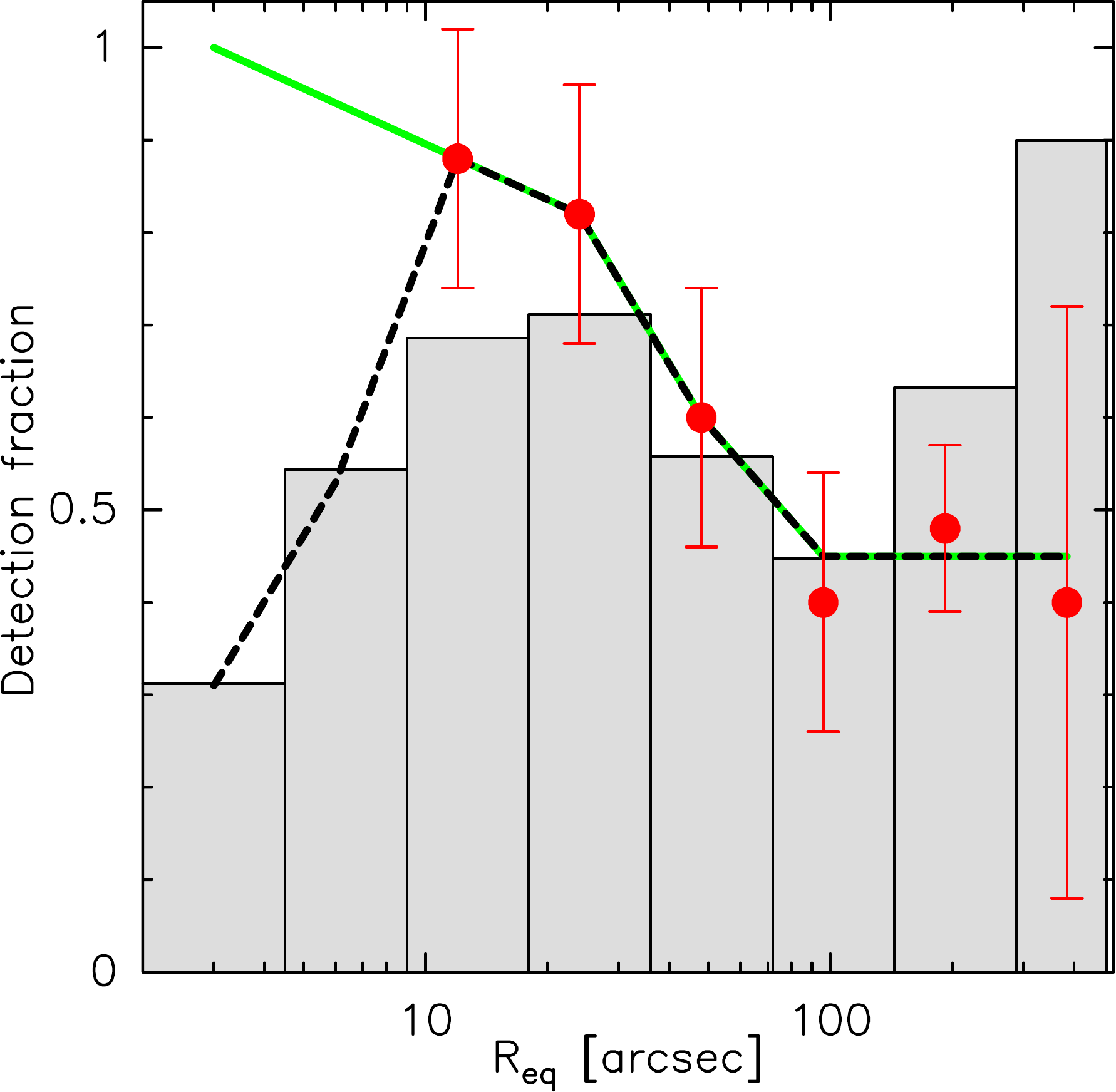}
\caption{Detection fraction as a function of SDC angular radius.The grey histogram is the same in the right-hand-side panel of Fig.~\ref{detect}. The red symbols are the same as in the left-hand-side panel of Fig.~\ref{visu}. The black dashed and green solid lines represent the two assumptions that have been made for the fraction of real SDCs (Sec. 5.2). 
} 
\label{fraction_final}
\end{figure}

\subsection{The fraction of real and spurious SDCs}

The main unknown in determining the overall reliability of the SDCs is the detailed behaviour of the fraction of real clouds at small sizes which \herschel\ cannot resolve. For larger sources, a good estimate of the fraction of real SDCs as a function of size is provided by the visual detection fraction extrapolated to the entire sample. For the two smallest bins, the real SDC fraction remains unknown. However, given that the trend shows an increase of that fraction with decreasing sizes (as expected from the simulations) one could argue that the real SDC fraction must keep increasing for the two smallest size bins. The other extreme assumption on can make is that the real SDC fraction at these sizes is given by the automated detection fraction. These two hypothesis are represented in Fig.~\ref{fraction_final} as green solid and black dashed lines. 
Taking the average of these two assumptions, one can now estimate the integrated real fraction of clouds over the entire sample. The fraction of real SDCs is estimated to be $76(\pm19)\%$, however for clouds with angular radius above 32\arcsec, this real SDC fraction goes down to $\sim55(\pm12)$\%.  This decrease in the fraction of real clouds at large sizes is a reflection of the increasing fraction of artefacts in the 8\microns\ background images (see Sec.~5.1). The quoted uncertainties result from the combination of: a 3\% uncertainty related to the two different assumptions regarding the fraction of real small SDCs (see above); a 14\% Poisson uncertainty per size bin related to the small number statistics of the visual inspection, this uncertainty goes down to 6\% when considering all 6 size bins, and to 7\% when consider only the 4 largest size bins; a systematic error of 10\%  related to the visual real/spurious classification, this error goes down to 5\% when only considering the largest clouds (it is easier to visually characterize the nature of larger clouds).

It is worth noting that using SCUBA 850\microns\ data, \citet{parsons2009} estimated that 75\% of the 205 {\it MSX} IRDCs from the \citet{simon2006a}  catalogue they analysed were real. This percentage is in total agreement with the detection fraction we provide here for the \spitzer\ IRDCs.

For comparison, we checked on a one to one basis our detection results with the one from \citet{wilcock2012} for the $l=[300\degr - 330\degr]$ region. Where they identified 38\% of the SDCs of this region as being \herschel\ bright, we detect 61\%, only marginally less than the average over the entire sample. Of these 38\% identified by Wilcock et al. (2012), 82\% are also identified as real by our detection scheme. Of the remaining 18\%, 72\% have angular radius smaller than 16\arcsec, corresponding to the size bins for which our identification scheme is less complete (see Fig.~\ref{fraction_final}).

The clouds we positively identified but which were missed by Wilcock et al. (2012), i.e. 23\% of the cloud population in the $l=[300\degr - 330\degr]$ region, are mostly low column density clouds that appear to be faint at 250\microns. This explains why, based on a visual inspection at that wavelength, they were missed. An example of such cloud is shown in Fig.~\ref{irdc}.

\begin{figure}[t]
\centering
\includegraphics[width=9.cm]{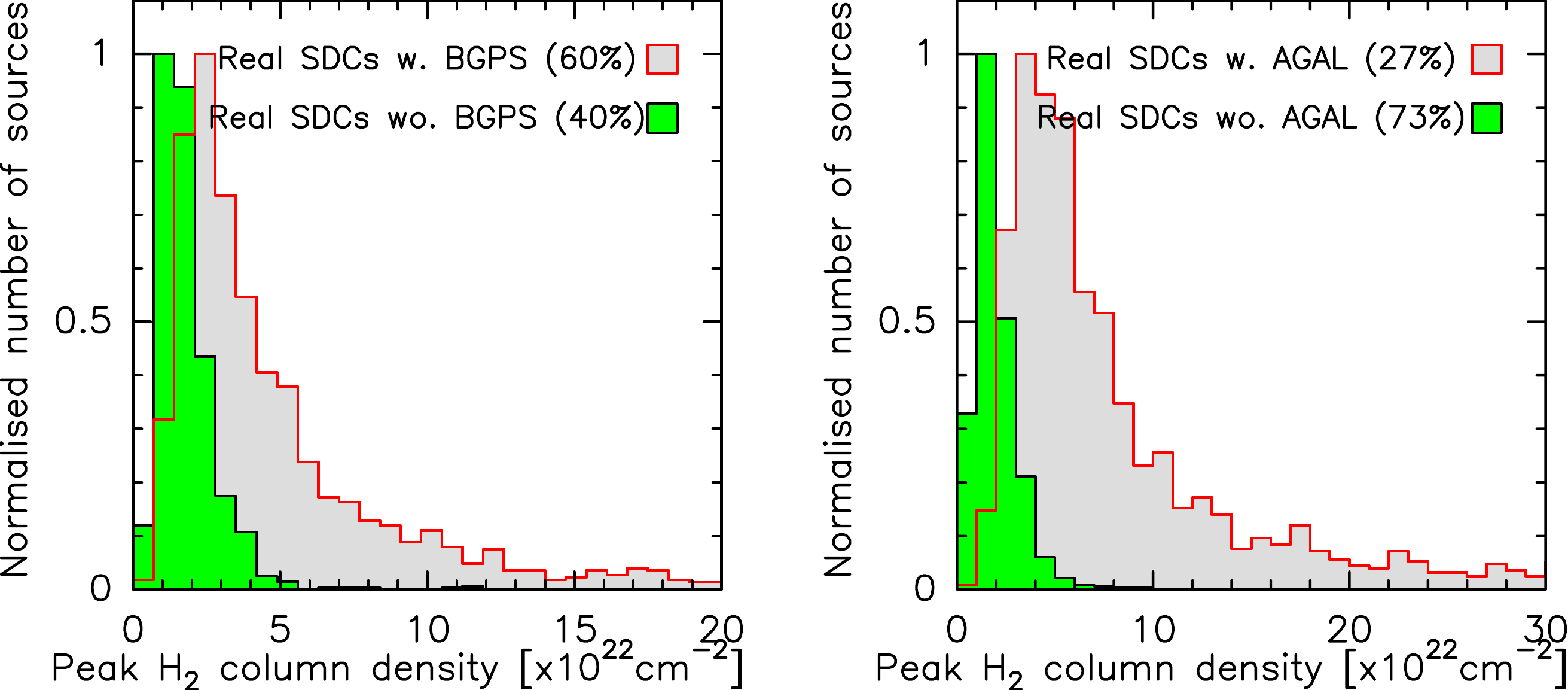}
\caption{ (Left): Herschel H$_2$ peak column density histogram of \herschel-confirmed SDCs with a BGPS counterpart (grey histogram) and without a BGPS counterpart (green histogram). (Right): Same as the left-hand-side histogram but for ATLASGAL counterparts. The percentage of \herschel-confirmed SDCs in each category is indicated in the top right corner of each panel.
} 
\label{comp_bgps_agal}
\end{figure}

\section{Spitzer dark clouds in the BGPS and ATLASGAL}

 In order to further characterise {\it Spitzer} dark clouds, we crosschecked the \herschel-confirmed SDC sample against the catalogues of both BGPS \citep{rosolowsky2010} and ATLASGAL \citep{csengeri2014} (sub-)millimetre surveys. Differences in angular resolution, data type (emission versus extinction), and shapes of these sources make the association difficult to define. We considered that there was {\it association} between a SDC source and a BGPS/ATLASGAL source when the distance between the centroid positions of the two sources was less than the sum of their radii. The SDC radii are provided in Table 1 of this paper. For the BGPS sources we used the values quoted in column 10 of Table 1 of \cite{rosolowsky2010}. For ATLASGAL sources, we used, as source radius, the values quoted in column 8 of Table 1 of \cite{csengeri2014}. Figure \ref{comp_bgps_agal} shows histograms of peak H$_2$ column density (see Table 1) for \herschel-confirmed SDCs with and without BGPS counterparts (left panel) and real SDC with and without ATLASGAL counterparts (right panel). On this figure we can see that 1,408 of the 2,333  (60\%) of the real SDCs covered by the BGPS have a BGPS counterparts while 925 (40\%) do not. On this histogram it is clear that the latter represent the lowest column density (and smallest) SDCs of the catalogue. These are missed by BGPS as a result of their small sizes and their corresponding small BGPS beam filling factor. On the right panel of Fig.~\ref{comp_bgps_agal}, one can see that only 1,907 (27\%) of the real SDCs covered by ATLASGAL have an ATLASGAL counterpart. This fraction is in good agreement with \citet{contreras2013} who find an association fraction of 30\%. Here again, SDCs without an ATLASGAL counterpart are mostly at low column density. The reason for which the percentage of SDCs with BGPS sources is larger is due to the difference in source identification schemes used in BGPS and ATLASGAL. The latter focused the source identification on rather compact (upper limit of 50\arcsec) and centrally concentrated (as imposed by the Gaussian fitting routine) sources. Such constraints are not imposed in the BGPS extraction. This comparison shows there is a rather large population of cold and compact sources that are missed by current (sub-)millimetre galactic plane surveys.

Note also that a large number of BGPS and ATLASGAL sources do not have SDC counterparts. While the majority of these sources are infrared bright sources, and therefore cannot be associated, by definition, with an infrared dark cloud, a fraction of them are infrared dark sources that have not been included in the the PF09 catalogue.  \cite{ellsworth2013} identified a sub-sample of infrared dark BGPS sources and crosschecked against the SDCs from PF09. In their study,  \cite{ellsworth2013} find that  $\sim70\%$ (based on their Fig.~15) of the sources from their sample are low contrast IR dark sources that remained undetected by PF09. In their paper,  \cite{ellsworth2013} considered two sources to be associated if the distance between the centroid of the two sources is smaller than the semi-major axis of the SDC source. This is a very restrictive association condition for two main reasons. The first reason is linked to the definition of the semi-major axis. The semi-major axis $\sigma_{maj}$ of a SDC  is defined as the column density weighed distance {\it dispersion} from the centroid position in the direction of the source major axis. Therefore, the disc of area $\pi \sigma_{maj}^2$ will have a much smaller area than $\pi R_{eq}^2$ where $R_{eq}$ is the radius of the disc of the a same area as the source, and BGPS sources outside the disc of radius $\sigma_{maj}$ will be missed. The second reason is linked to the shape of SDCs. Sources with elongated/complicated shapes will have a large fraction of their area beyond the association radius (even when considering the $ R_{eq}$ as the association radius) and very elongated filaments, for instance, having BGPS sources at their tips will be missed. This is exactly what happened for the source \cite{ellsworth2013} use for illustration in their paper (BGPS \# 5647) and for which they claim that the lower part of the cloud is not part of the PF09 catalogue while it actually is (this can be seen by looking at the image of SDC35.527-0.269 on www.irdarkclouds.org). 

\begin{figure}[t]
\centering
\includegraphics[width=9.cm]{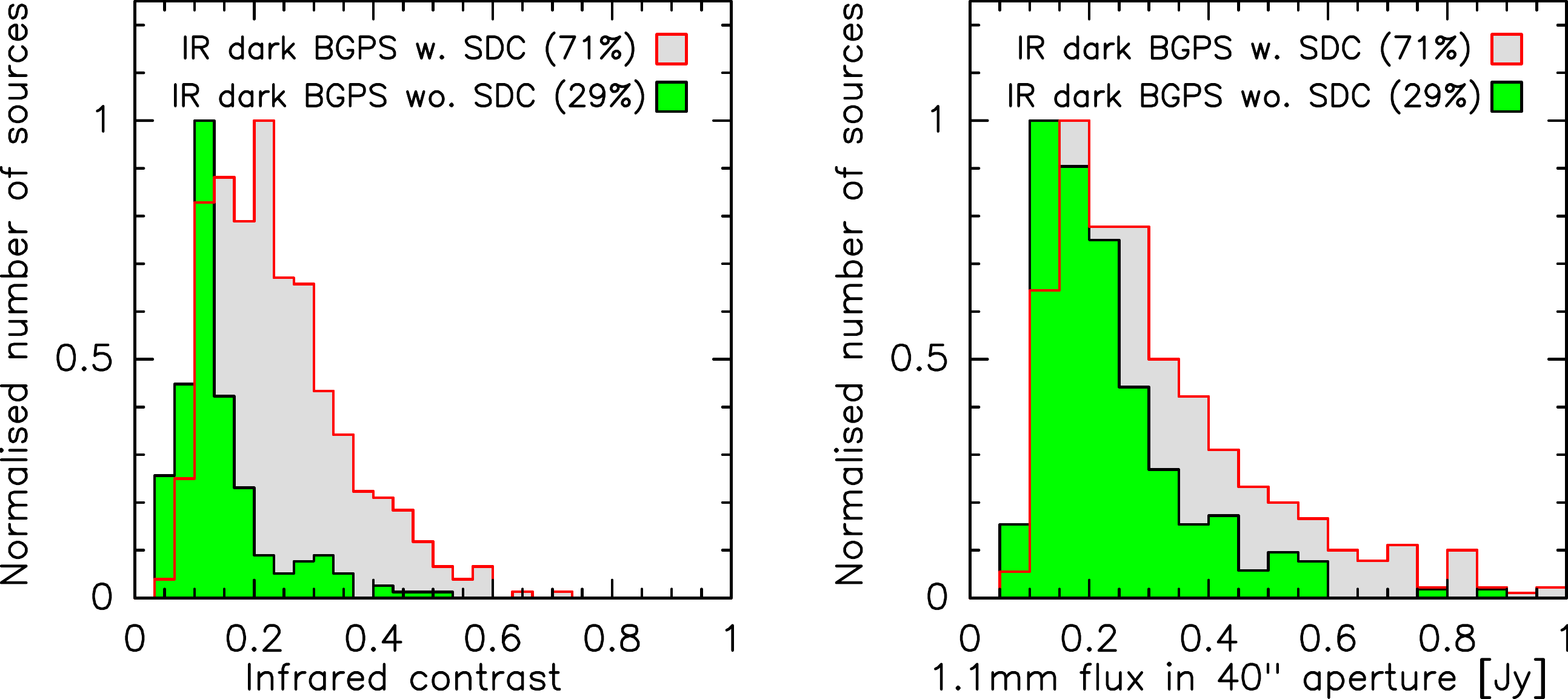}
\caption{  (Left): Distribution of infrared contrast for BGPS sources associated with \herschel-confirmed SDCs (grey) and BGPS sources without SDC association (green). (Right): Same as in the left-hand-side panel but showing the distribution of BGPS 1.1mm flux as estimated within a 40\arcsec\ aperture radius. The percentage of BGPS sources in each category is indicated in the top right corner of each panel.
} 
\label{comp_bgps_sdc}
\end{figure}

We therefore performed the association of IR dark BGPS sources with \herschel-confirmed SDC sources using the same association condition as we used previously, and built the histograms of infrared contrast (see equation 11 of \cite{ellsworth2013}) and BGPS 1.1mm 40\arcsec\ aperture flux. Figure \ref{comp_bgps_sdc} displays these histograms. First, we see that, with our association condition, only 29\% of BGPS sources are not associated to a SDC as opposed to 70\% in \cite{ellsworth2013} analysis. In the left panel of Fig.~\ref{comp_bgps_sdc} we see that this population of sources are mostly low infrared contrast sources, as already noted by \cite{ellsworth2013}. However, the corresponding distribution is much more peaked (see Fig.~15 of their paper). Looking at images of individual sources with contrast above 0.2, we also notice that these BGPS sources are, in fact, associated with SDCs. The reason for which we failed to associate these BGPS sources with SDCs is the same one as that mentioned above: an elongated  BGPS source that includes a SDC at its tip can fail to pass the association criterion as the distance between the centroids of both sources can be larger that the sum of their $R_{eq}$ radii. These same relatively high contrast sources are also the ones making the high-end tail of the 1.1mm flux distribution in the right hand side panel of Fig.~\ref{comp_bgps_sdc}. 

In order to determine the nature of the remaining low contrast infrared dark BGPS sources without SDC association, we looked at both their BGPS and 8\microns\ Spitzer images. These sources appear to be mostly low column density IRDCs, as suggested by the position of the peak of 1.1mm flux distribution, with large beam filling factor (as opposed to the population of low column density SDCs undetected in BGPS data - see green histogram in Fig.~\ref{comp_bgps_agal}). These sources are either isolated sources or lying in the low density outskirts of denser clumps.

\section{Summary and conclusion}

Using \herschel\ Hi-GAL data we constructed H$_2$ column density images of the Galactic plane at 18\arcsec\ resolution using the 160\microns/250\microns\ ratio as a probe of the dust temperature. We used these data to determine the fraction of real IRDCs from the Peretto \& Fuller (2009) catalogue by analysing their \herschel\ column density properties. Simulating the detection process, along with performing a visual inspection of a small sub-sample of SDCs,  show that small angular size clouds are missed by our automated identification scheme as a result of beam dilution and large background fluctuations. On the other hand, very large  features can be wrongly identified as real clouds due to the probability of finding an \herschel\ column density peak in a given area of the Galactic plane. Taking these effects into account, we estimated that $76(\pm19)\%$ of the SDCs are real. This fraction decreases to $\sim55(\pm12)\%$ when considering clouds with an angular radius larger than $\sim30$\arcsec.

The availability of \herschel\  data towards the sources of the PF09 catalogue gives us the opportunity to analyse their far-infrared counterparts. As a result, the SDC properties are much better constrained, and studies of the earliest stages of Galactic star formation can now be more reliably performed, both on an individual and global scale. One particularly interesting feature of this \herschel-confirmed SDC sample is the broad range of cloud sizes/column densities it probes, providing a unique opportunity to study the link between the earliest stages of low- and high-mass star formation across the Milky Way.

\begin{acknowledgements}
We thank the anonymous referee whose report helped improve the quality of this paper. NP acknowledges the support from the UK Science \& Technology Facility Council (STFC) via grant ST/M000893/1. CL  has been funded via a STFC PhD studentship. GAF and AT acknowledge the support from STFC via grant ST/J001562/1. MAT  acknowledges support from STFC via grant ST/M001008/1.
\end{acknowledgements}

\bibliographystyle{aa}
\bibliography{references}

\begin{appendix}
\section{A sample of randomly selected SDCs }
\begin{figure*}[t]
\centering
\includegraphics[width=16.5cm]{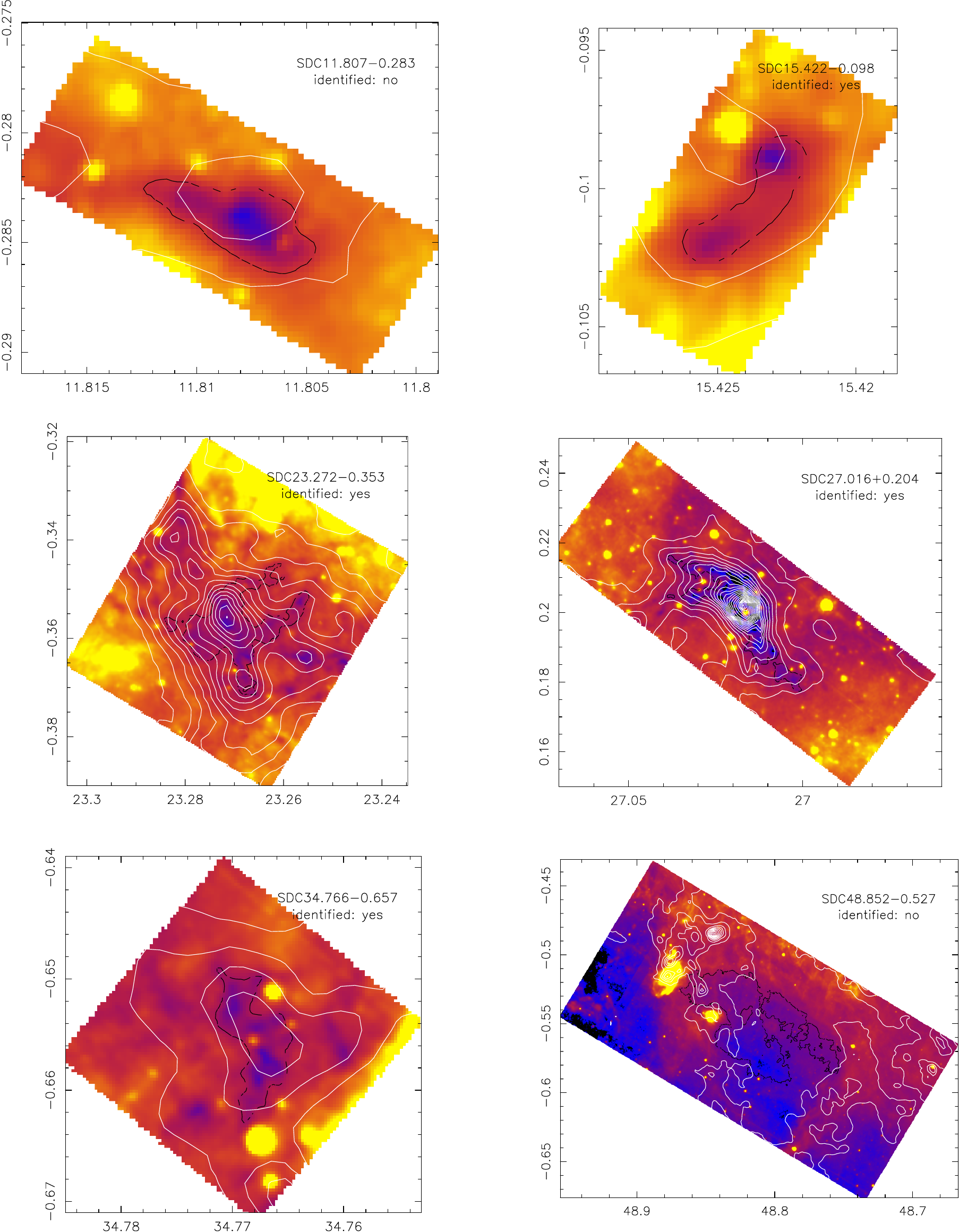}
\caption{Images of 6 randomly selected SDCs. The colour scale is the \spitzer\ 8\microns\ emission. The black contour is the $\tau_{8\mu m}=0.35$ contour marking the boundary of the SDC as originally identified in PF09. The white contours are the \herschel\ H$_2$ column density contours, all starting at $0.1\times10^{22}$~cm$^{-2}$, and separated by  $0.5\times10^{22}$~cm$^{-2}$. The axes of the images are galactic coordinates in degrees. In the top right corner we give the name of the SDC and if our identification scheme has recognised them as being identified with a \herschel\ column density peak. Note that SDC11.807-0.283 is not identified while, by eye, it seems clearly associated with a faint peak. Criterion $c_2$ for this cloud is slightly below our threshold value of 3, explaining why it is not picked up by our identification scheme.
} 
\label{app1}
\end{figure*}

\begin{figure*}[t]
\centering
\includegraphics[width=16.cm]{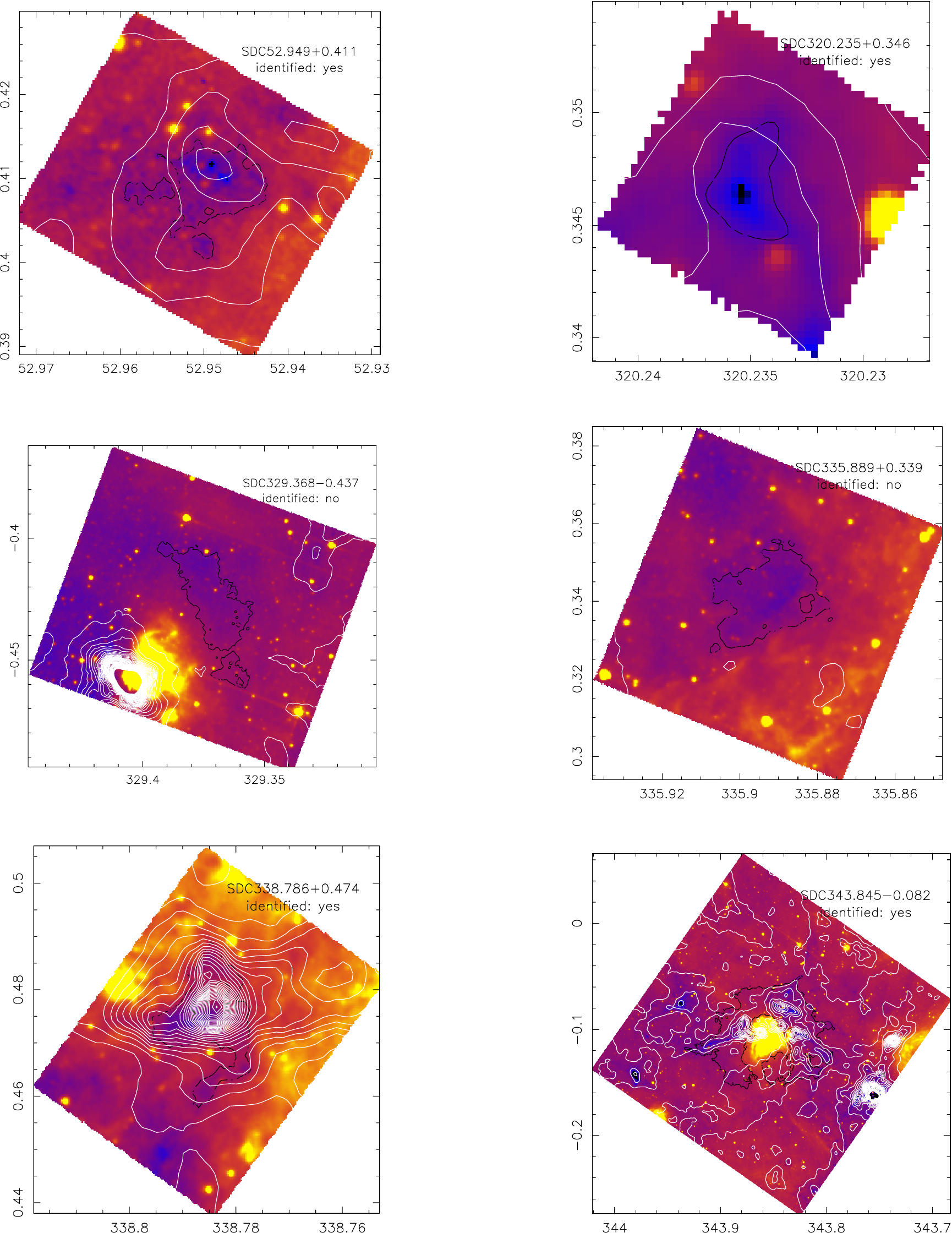}
\caption{Same as Fig.~\ref{app1}. The \herschel\ contours for SDC343.845-0.082 have been spaced by $1\times10^{22}$~cm$^{-2}$ (as opposed to $0.5\times10^{22}$~cm$^{-2}$ for the others) for a matter of clarity.} 
\label{app2}
\end{figure*}

\end{appendix}

\end{document}